\documentclass[3p,times,procedia]{elsarticle}
\usepackage{nupha_ecrc}
\usepackage{amsmath}
\usepackage{amssymb}
\usepackage{amsfonts}
\usepackage{mathrsfs}  
\usepackage{bm}
\usepackage[utf8]{inputenc}

\volume{00}

\firstpage{1}

\journalname{Nuclear Physics A}

\runauth{}

\jid{nupha}

\jnltitlelogo{Nuclear Physics A}

\usepackage{amssymb}

\usepackage[figuresright]{rotating}

\begin{document}

\begin{frontmatter}

\dochead{XXVIIIth International Conference on Ultrarelativistic Nucleus-Nucleus Collisions\\ (Quark Matter 2019)}

\title{Momentum-dependent flow fluctuations as a hydrodynamic response to initial geometry}

\author[unicamp]{M.~Hippert}

\author[unicamp]{D.D.~Chinellato}

\author[usp]{M.~Luzum}

\author[uiuc]{J.~Noronha}

\author[ufsc]{T.~Nunes da Silva}

\author[unicamp]{J.~Takahashi}

\address[unicamp]{Instituto de F\'isica Gleb Wataghin, Universidade Estadual de Campinas, Rua S\'ergio Buarque de Holanda 777, 13083-859 S\~ao Paulo, Brazil}
\address[usp]{Instituto de F\'{\i}sica, Universidade de  S\~{a}o Paulo,  Rua  do  Mat\~{a}o, 1371,  Butant\~{a},  05508-090,  S\~{a}o  Paulo,  Brazil}
\address[uiuc]{Department of Physics, University of Illinois, 1110 W. Green St., Urbana IL 61801-3080, USA}
\address[ufsc]{Departamento de  F\'{\i}sica - Centro de Ci\^encias  F\'{\i}sicas e Matem\'aticas, Universidade Federal de Santa Catarina, Campus Universit\'ario Reitor Jo\~ao David Ferreira Lima, Florian\'opolis 88040-900, Brazil}

\begin{abstract}
We propose a redefinition of the principal component analysis (PCA) of anisotropic flow that makes it more directly connected to fluctuations of 
the initial geometry of the system. 
Then, using state-of-the-art hydrodynamic simulations, we make an explicit connection between flow fluctuations and a cumulant expansion of  
the initial transverse geometry. %
In particular, we show that the second principal component of elliptic flow is generated by higher-order cumulants, and therefore probes smaller length scales of the initial state. 
With this information, it will be possible to put new constraints on properties of the early-time dynamics of a heavy-ion collision, including small-scale structure, 
as well as properties of the quark-gluon plasma.
\end{abstract}

\begin{keyword}
 heavy-ion collisions \sep collective dynamics \sep anisotropic flow \sep principal component analysis
\end{keyword}

\end{frontmatter}

\section{Introduction}
\label{sec:intro}

In ultrarelativistic heavy-ion collisions, momentum-space anisotropies in the final distribution of particles are understood to originate from 
pressure gradients in the early stages of the quark-gluon plasma (QGP).  
Here, we study how the small-scale structure of the initial state affects momentum-dependent correlations of the anisotropic flow, 
which can be efficiently visualized by employing a principal component analysis (PCA) \cite{Bhalerao:2014mua,Hippert:2019swu}. 
In order to connect this analysis with features of the initial state, we 
redefine the PCA observables and 
investigate how cumulants of the initial geometry 
correlate with flow harmonics in the final state, in an event-by-event basis \cite{Teaney:2010vd,Gardim:2011xv,Gardim:2014tya}.

\section{Principal Components of Anisotropic Flow}
\label{sec:PCA}

Principal component analysis is a statistical method that allows one to find, among a set of correlated variables, linearly independent combinations of 
maximum variance. 
Here, we study the PCA of the flow harmonics $V_n(p_T)$, defined as Fourier coefficients of the azimuthal distribution of particles 
at transverse momentum $p_T$:
\begin{equation}
 \dfrac{d N}{ p_T d p_T \,d\varphi} =\dfrac{1}{2\pi} %
 N(p_T) \sum_{n=-\infty}^{\infty} { {V}_n(p_T)}\,e^{- i n \varphi}\,.
 \label{eq:FourierYield}
\end{equation}  
The principal components $ V^{(\alpha)}_n({p_T})$ are found by diagonalizing the flow covariance matrix, such that:
\begin{equation}
{V}_{n\Delta}({p_T}_1,{p_T}_2) = %
\langle V_n({p_T}_1)\,V^*_n({p_T}_2) \rangle
\simeq \sum_{\alpha=1}^{\alpha_{\textrm{max}}}  V^{(\alpha)}_n({p_T}_1)  V^{(\alpha)}_n({p_T}_2)\,,
\label{eq:PCA}
\end{equation}
where %
 the components 
are put in descending order: $||V^{(\alpha)}_n|| \geq ||V^{(\alpha+1)}_n||$. 
Since a large hierarchy is found among the different components,  
the RHS of Eq.~\eqref{eq:PCA} might be truncated, say, at $\alpha_{\textrm{max}}=2$ \cite{Bhalerao:2014mua}. 

In \cite{Hippert:2019swu}, we have shown that the original flow PCA observables proposed by Bhalerao \textit{et al.}\ in \cite{Bhalerao:2014mua}, 
and measured by the CMS Collaboration \cite{Sirunyan:2017gyb},  
are sensitive not only to fluctuations of anisotropic flow $V_n(p_T)$, but also to fluctuations of the number of particles in each bin, $N({p_T})$. 
The reason behind this is that they were defined from the covariance matrix 
 ${V}_{n\Delta}^N({p_T}_1,{p_T}_2)%
= \langle { N({p_T}_1)}\,{N({p_T}_2)}\,V_n({p_T}_1)\,V^*_n({p_T}_2) \rangle$. 
Since the anisotropic flow per particle $V_n$ is better correlated with the initial geometry, the sensitivity to fluctuations of 
$N({p_T})$ is undesirable for our purposes.\footnote{While fluctuations of particle number can be of interest, they can be 
characterized separately by the covariance matrix $\langle \Delta N({p_T}_1) \Delta N({p_T}_2)\rangle$ \cite{Bhalerao:2014mua,Gardim:2019iah}.} 
To solve this issue, we divide the covariance matrix $V_{n\Delta}^N$ %
 by $\langle N({p_T^a})  N({p_T^b})\rangle$, 
so as to approximately cancel out particle-number fluctuations \cite{Hippert:2019swu}: 
\begin{equation}
 {V}_{n\Delta}^R({p_T}_1,{p_T}_2)%
= \dfrac{\langle { N({p_T}_1)}\,{N({p_T}_2)}\,V_n({p_T}_1)\,V^*_n({p_T}_2) \rangle}{\langle  N({p_T}_1)   N({p_T}_2)\rangle}
\simeq \langle V_n({p_T}_1)\,V^*_n({p_T}_2) \rangle\,. 
\label{eq:PCAExT}
\end{equation}

\begin{figure}
\centering
 \includegraphics[width=\textwidth]{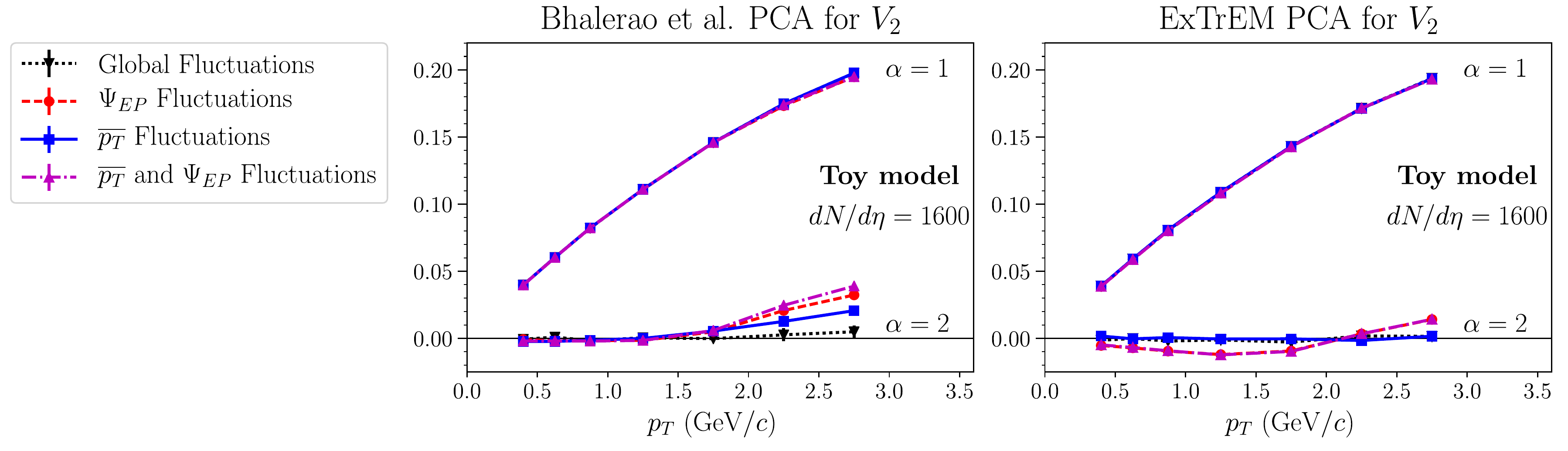} 
\caption{Toy model results for the PCA of elliptic flow for the original definition in Bhalerao \textit{et al.} \cite{Bhalerao:2014mua} (left panel) and the new improved definition 
proposed in \cite{Hippert:2019swu} (right panel). 
Scenario 1 (black dotted curve) includes only global fluctuations; 
scenario 2 (red dashed curve) also includes subleading flow in the form of event plane fluctuations; scenario 3 (solid blue curve) includes 
 fluctuations of the mean transverse momentum; scenario 4 (dot-dashed magenta curve) includes all the three sources of fluctuations. 
 Both definitions of the PCA yield a subleading mode in the presence of subleading flow, and neither of them are sensitive to global multiplicity fluctuations. 
 However, the Bhalerao \textit{et al.}\ PCA \cite{Bhalerao:2014mua} is sensitive to fluctuations of the mean transverse momentum, while the new, improved PCA is 
 sensitive to subleading flow alone \cite{Hippert:2019swu}. 
}
\label{fig:ToyModel}
\end{figure}

To elucidate how the different PCA observables respond to particle-number and anisotropic flow fluctuations, 
we study them in a controlled toy model. We sample the distribution of particles in $p_T$ with 
an azimuthal distribution with constant $|V_2(p_T)|$ and a simple exponential spectra $N(p_T) \propto e^{-p_T/\overline{p_T}}$, with a stochastic value of the total charged multiplicity. 
We test the PCA observables within the toy model in four different scenarios, with
\begin{enumerate}
 \item only global particle-number fluctuations, with no transverse-momentum dependence; 
  \item {fluctuations} of the { event plane}, with some profile $\Psi_{EP}(p_T)$, as a simplified model of subleading flow; 
 \item {fluctuations} of the { spectrum}, with $\overline{p_T} \to \overline{p_T} + \delta \overline{p_T}$;
 \item fluctuations of both the event plane $\Psi_{EP}(p_T)$ and the mean transverse momentum $\overline{p_T}$. 
\end{enumerate}
 Results are shown in Fig.~\ref{fig:ToyModel}. 
   In this figure, we see that, unlike the original PCA observables proposed by Bhalerao \textit{et al.},  the new PCA observables are 
   insensitive to particle-number fluctuations, thus more directly probing fluctuations of the initial geometry. 

For more details on the redefinition of the PCA of anisotropic flow, see  Ref.~\cite{Hippert:2019swu}.

\section{Mapping Hydrodynamic Response}
\label{sec:Mapping}

In order to investigate the physical content of the principal components of anisotropic flow \cite{Mazeliauskas:2015vea,Mazeliauskas:2015efa}, we apply the 
mapping of hydrodynamic response to initial geometry from \cite{Teaney:2010vd,Gardim:2011xv,Gardim:2014tya}.  
This mapping provides an extension of the usual eccentricity scaling of elliptic flow through a systematic expansion, 
starting from the assumption that the posterior evolution of the system and the final one-particle distribution are 
fully determined by the transverse energy-density profile $\rho(\tau_0,\vec x)$ at an initial proper time $\tau =\tau_0$.
The initial transverse profile of the system, in turn, can be fully characterized by a (infinite) set of cumulants $W_{n,m}$ \cite{Gardim:2014tya}. 
For instance,
\begin{equation}
W_{0,2} = \frac{i^2}{4}\,\bigg[\{z^* z\} -  \{z^*\}\{z\} \bigg]\,,
\qquad
 W_{2,2} = \frac{i^2}{8}\,\bigg[\{z^2\} -  \{z\}^2 \bigg]\,,
 \qquad  \{ (\cdots)\} \equiv \frac{\int d^2 x\, \rho(\vec x) \,(\cdots)}{\int d^2 x \,\rho(\vec x)}\,,
\end{equation}
where $z\equiv x+iy$ and we have defined the ``spatial average'' $\{(\cdots)\}$. From them, we can define generalized eccentricities 
\begin{equation}
 \epsilon_{n,m} \equiv - W_{n,m}/(W_{0,2})^{m/2}\,,
\end{equation}
where the first index $n$ specifies the symmetry under rotations, while $m$ is related to the typical length scales being probed, 
with higher values of $m$ corresponding to increasingly finer details of the initial conditions.  

Since we assume that the full evolution of the system is determined from $\rho(\tau_0,\vec x)$, the flow harmonics $V_n$ can be predicted by the set of all
eccentricities and, 
if they are not too large, %
can be approximated by a Taylor series in $\epsilon_{n,m}$. 
Also assuming that large scales dominate the final flow pattern, we can truncate the expansion of $V_n[\{\epsilon_{n,m}\}]$ at a finite value of $m$. 
The elliptic flow $V_2$, for instance, is approximated by 
\begin{align}
\begin{split}
    V_2(p_T) \simeq \kappa_0(p_T)\,\epsilon_{2,2} & %
 +  \kappa_1(p_T)\,\epsilon_{2,4} +\kappa_2(p_T)\,\epsilon_{2,6}  +\kappa_3(p_T)\,\epsilon_{2,8} + \mathcal{O}(m=10)\\
 &+    \kappa_4(p_T)\,|\epsilon_{2,2}|^2\epsilon_{2,2} +   \kappa_5(p_T)\,\epsilon_{4}\epsilon_{2}^*  
  +   \kappa_6(p_T)\,\epsilon_{1,3}^2 + \ldots  
  + \mathcal{O}(\epsilon^3)\,,
  \label{eq:predV2}
\end{split}
\end{align}
where linear terms are placed on the first line and only terms with the correct symmetry under rotations are included. 
In hydrodynamic simulations, the coefficients $\kappa_i(p_T)$ can be independently fixed for each $p_T$ bin, 
by maximizing the correlation between the final $V_2(p_T)$ and the predictor 
in Eq.~\eqref{eq:predV2}.

\section{Results and Conclusions}
\label{sec:Results}

\begin{figure}
\centering
 \includegraphics[width=\textwidth]{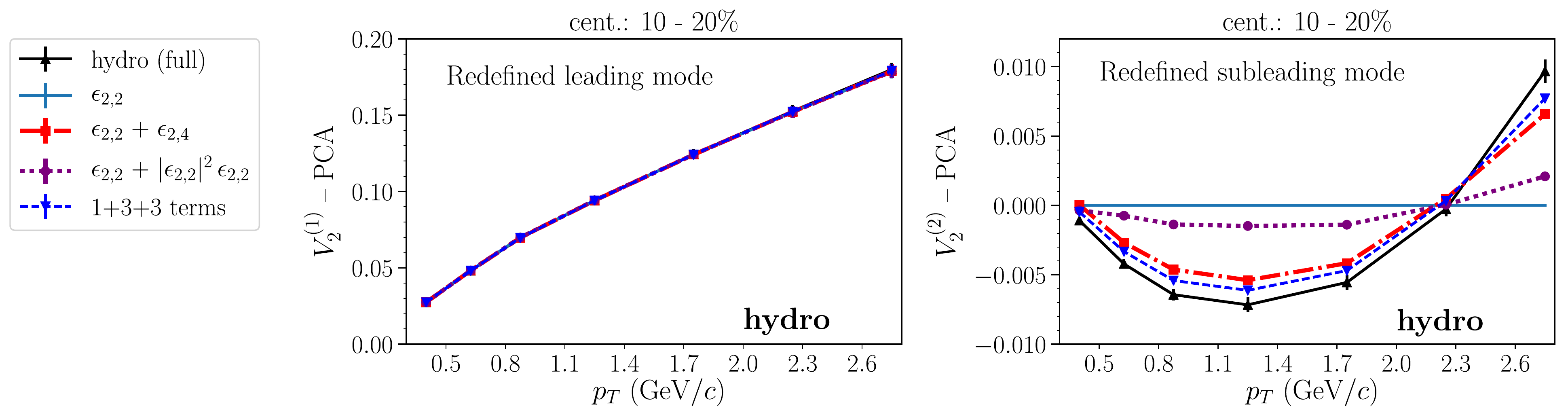} 
\caption{Comparison between the principal components of the elliptic flow per particle in a full hydrodynamic simulation (solid black curve) and the ones predicted from Eq.~\eqref{eq:predV2} (dashed blue curve).
Also shown are the predictions from Eq.~\eqref{eq:predV2} when only the first subleading linear term $\epsilon_{2,4}$ (dash-dotted red curve) or the leading nonlinear term $|\epsilon_{2,2}|^2\epsilon_{2,2}$ (magenta dotted line) are considered. 
While the leading principal component (left panel) is correctly predicted as long as the leading eccentricity $\epsilon_{2,2}$ is included, the subleading principal component (right panel) is only reproduced if the linear response to $\epsilon_{2.4}$ is included.}
\label{fig:PCAmapping}
\end{figure}

To investigate the connection between the  PCA of the flow per particle 
and the initial geometry, we employ a  state-of-the-art event-by-event hydrodynamic model (TrENTO+MUSIC+iSS+UrQMD), 
with parameters optimized to reproduce LHC data  
\cite{Hippert:2019swu,NunesdaSilva:2018viu,Schenke:2010nt,Schenke:2011bn,Bass:1998ca,Bleicher:1999xi,Moreland:2014oya,Bernhard:2018hnz}. 
In particular, we test if the PCA of the flow per particle can be reproduced by the approximation in  Eq.~\eqref{eq:predV2}, 
and study which terms in this expression are the most important for each principal component. 
Results are shown in Fig.~\ref{fig:PCAmapping}, where we 
check that the description of flow fluctuations is visibly improved by considering higher-order eccentricities of the initial geometry. 
In particular, we find that the second principal component of $V_2$ is largely sensitive to the subleading eccentricity $\epsilon_{2,4}$,  
which is sensitive to the smaller-scale structure of the initial geometry. 
In this figure, the PCA is defined directly from $\langle V_n ({p_T}_1) V_n^*({p_T}_2) \rangle$, which is approximately equivalent to 
the matrix in Eq.~\eqref{eq:PCAExT}  \cite{Hippert:2019swu}. 

We conclude that the redefined subleading principal component of elliptic flow is a valuable probe of fluctuations of 
the initial stages at smaller length scales. 
As such, it will provide fresh insight into the formation and properties of the QGP, 
establishing new constraints for models of high-energy heavy-ion collisions.

\section*{Acknowledgments}

This research was funded by FAPESP grants number 2016/13803-2 (D.D.C.),
2016/24029-6 (M.L.), 2017/05685-2 (all), 2018/01245-0 (T.N.dS.) and 2018/07833-1(M.H.). 
D.D.C., M.L., J.N., and J.T. thank CNPq for financial support.
We also acknowledge computing time provided by the Research Computing
Support Group at Rice University through agreement with the University of São Paulo.

\bibliographystyle{elsarticle-num}
\bibliography{mappingPCA}

\end{document}